# Coexistence of superconductivity and magnetism in $K_{0.8}Fe_2Se_{1.4}S_{0.4}$

L. Li,[1] Z. R. Yang,[2*] Z. T. Zhang,[1] W. Tong,[3] C. J. Zhang,[3] S. Tan,[1] and Y. H. Zhang[1,3]

[1]High Magnetic Field Laboratory, University of Science and Technology of China, Hefei 230026, People's Republic of China

[2]Key Laboratory of Materials Physics, Institute of Solid State Physics, Chinese Academy of Sciences, Hefei 230031, People's Republic of China

[3]High Magnetic Field Laboratory, Hefei Institutes of Physical Science, Chinese Academy of Sciences, Hefei 230031, People's Republic of China

**High-quality single crystals of $K_{0.8}Fe_2Se_{1.4}S_{0.4}$ are successfully synthesized by self-flux method with the superconducting transition temperatures $T_c^{onset}$ = 32.8 K and $T_c^{zero}$ = 31.2 K. In contrast to external pressure effect on superconductivity, the substitution of S for Se does not suppress $T_c$, which suggests that chemical doping may mainly modulate the anion height from Fe-layer rather than compressing interlayer distance. The investigation of the micromagnetism by electron spin resonance shows clear evidence for strong spin fluctuation at temperatures above $T_c$. Accompanied by the superconducting feature spectra, a novel resonance signal develops gradually upon cooling below $T_c$, indicating the coexistence of superconductivity and magnetism in $K_{0.8}Fe_2Se_{1.4}S_{0.4}$ crystal.**

During the past few years, a lot of research efforts have been devoted to iron-based superconductors, mainly focusing on the interplay between superconductivity and magnetism.[1-11] The parent compounds of LnFeAsO (Ln is rare earth element),[1-5] AeFe$_2$As$_2$ (Ae is alkali earth element)[6-9] and iron chalcogenides FeSe(Te)[10-11] exhibit spin-density-wave (SDW) or antiferromagnetic (AF) order. Chemical doping suppresses the magnetism and superconductivity appears at a certain doping value.[12] In many cases, the superconductivity is found to coexist with magnetism either in atomic scale or in nanoscale with electronic phase separation characteristics.[11-13] All these observations clearly indicate that the superconductivity in the iron-based superconductors is proximity to magnetism and spin fluctuation may play an important role for superconducting paring.[12] A direct correlation of superconductivity to spin fluctuation is the pressure effect on superconducting transition temperature $T_c$ of FeSe.[14-15] $T_c$ being dramatically increased to 37 K under high pressures is accompanied by the enhancement of spin fluctuation.[14] For higher pressures, $T_c$ decreases due to structure transition.[15] It is conjectured that if the structure transition under high pressure could be avoided, higher $T_c$ can be expected in the iron chalcogenides.

Very recently, superconductivity with $T_c$ around 30 K has been reported in FeSe layered compound A$_x$Fe$_2$Se$_2$ (A = K, Rb, Cs, Tl) with intercalating the alkaline metals between the FeSe layers.[16-21] The first-principles electronic structure calculations indicated that the ground state of AFe$_2$Se$_2$ is quasi-2-dimensional bi-collinear or stripe-like antiferromagnetic with a magnetic moment of 2.8 or 2.26 $\mu_B$ around each Fe atom.[22-23] μSR measurements on $Cs_{0.8}(FeSe_{0.98})_2$ showed that the superconducting state microscopically coexists with a rather strong magnetic phase.[24] NMR experiments revealed weak FM or AFM fluctuations in $K_xFe_2Se_2$.[25-26] Evolution from a superconducting state to an AFM insulating state has been reported by varying Fe content.[21,27] These recent discoveries allow to perform a direct comparison between FeSe(Te) and $AFe_2Se_2$. In FeSe(Te) system, the interplay between magnetism and superconductivity is mainly investigated by changing the ratio of anion.[10-13] The first-principles calculations indicated that it is not chalcogen species but anion height from the Fe-plane that determines the magnetism.[28] Therefore the study of anion substitution effect on superconductivity in $AFe_2Se_2$ may shed light on the mechanism of unconventional superconductivity in this system.

In this work, we successfully grew the single crystals of a new superconductor $K_{0.8}Fe_2Se_{1.4}S_{0.4}$ by self-flux method. The onset and zero-resistivity transition temperature were estimated to be 32.8 K and 31.2 K, respectively. The investigation of the micromagnetism by Electron spin resonance (ESR) showed clear evidence for strong spin fluctuation at temperatures above $T_c$. With decreasing temperature below $T_c$, a novel resonance signal develops gradually accompanied by the superconducting diamagnetic signal, suggesting the coexistence of superconductivity and magnetism.

Single crystals $K_{0.8}Fe_2Se_{1.4}S_{0.4}$ were grown from the melt using self-flux method. First, starting material $FeSe_{0.8}S_{0.2}$ was prepared with high-purity powder of iron, selenium and sulfur with Fe: Se: S= 1: 0.8: 0.2 at 650 for 12 hours. Then, K pieces and $FeSe_{0.8}S_{0.2}$ powder were put into a small quartz tube with nominal composition as $K_{0.8}Fe_2(Se_{0.8}S_{0.2})_2$. The small quartz tube was sealed under high vacuum and then was put in a bigger quartz tube following by evacuating and being sealed. The mixture was heated up to 1050 and kept over 4 hours. Afterwards the melt was cooled down to 700 with the cooling rate of 6 /h before the furnace was shut down.

The obtained crystals were characterized by powder X-ray diffraction (XRD) and X-ray single crystal diffraction with Cu $K_\alpha$ radiation at room temperature. Actual composition of the crystal was confirmed using energy dispersive x-ray spectrometry (EDXS). The resistivity was measured by using a standard four-probe method in a Quantum Design Physical Properties Measurement System (PPMS). Magnetic properties were investigated using a superconducting quantum interference device (SQUID) magnetometer. Electron spin resonance (ESR) measurements were performed in a Bruker EMX plus 10/12 CW-spectrometer at X-band frequencies ( $f$ 9.36 GHz) equipped with a continuous He gas-flow cryostat in the temperature region of 2–300 K by sweeping magnetic field parallel to ab-plane.

Figure 1 shows single crystal XRD patterns for the as-grown single crystals. Only (*00l*) diffraction peaks appear, indicating that the *c*-axis is perpendicular to the cleavage surface. To examine the purity of the prepared crystals, we further performed powder x-ray diffraction measurements by grinding partial crystals into powders. All the peaks can be well indexed with the symmetry of I4/mmm and no impurity phase is observed. The calculated lattice constants are c = 14.0344 Å and a = 3.8560 Å. Both are smaller than that reported in $K_{0.8}Fe_2Se_2$,[16-18] indicating the efficient substitution of S for Se in the as-grown single crystals. The actual composition of the crystals was estimated to be $K_{0.8}Fe_2Se_{1.4}S_{0.4}$ using an average of 6 points of the EDXS measurements.

The temperature dependence of resistivity for the $K_{0.8}Fe_2Se_{1.4}S_{0.4}$ single crystal is shown in Fig. 2. At high temperatures, resistivity increases with decreasing temperature and exhibits a broad hump around 130 K. With further cooling, metallic behavior is observed and superconducting transition appears at about 33 K. The abnormal hump in the resistivity curve has also been observed in $K_xFe_2Se_2$ and $Rb_{0.8}Fe_2Se_2$, seems to be a common feature of this system.[16, 18-20] The lower inset of Fig. 2 shows details of the superconducting transition. The onset and zero-resistivity temperature were estimated to be $T_c^{onset}$ = 32.8 K and $T_c^{zero}$ = 31.2 K, respectively. The transition width is about 1.6 K, suggesting the high quality and homogeneous nature of our sample. Resistivity as a function of temperature under the magnetic field (H = 14 T) applied in ab-plane and along c-axis is also shown in Fig. 2. The transition temperature of superconductivity is suppressed and the transition is broadened under external magnetic field, see the upper inset of Fig. 2. Obvious difference for the effect of field along different direction on the superconductivity indicates the existence of anisotropy in $K_{0.8}Fe_2Se_{1.4}S_{0.4}$ crystal. However, the hump peak in resistivity curve has no obvious shift when applying magnetic field no matter with H//c-axis or H//ab-plane.

Diamagnetism at low temperature can be clearly observed in magnetization measurements with magnetic field of 100 Oe, as seen in Fig. 3. The value of onset temperature $T_c^{mag}$ at about 31K from the magnetization measurements is almost identical to the superconducting transition temperature $T_c^{zero}$ in resistivity measurements. The inset of Fig. 3 shows magnetic susceptibility of $K_{0.8}Fe_2Se_{1.4}S_{0.4}$ with the magnetic field of 1 T applied parallel to the c-axis and ab-plane from 10 K to 300 K. At low temperature, superconducting trace still exists as revealed by the drop of susceptibility. Whether or not the external field is parallel to the c-axis, the magnetic susceptibility gradually decreases with increasing temperature, totally different from that reported in $A_xFe_2Se_2$ (A = Rb, Cs).[18-19] For $A_xFe_2Se_2$, the susceptibility was found to increases with temperature above a certain temperature in normal state. The difference suggests that the magnetism is changed by substituting S for Se. No anomaly was observed around 130 K where the resistance presents a hump.

Since S has a smaller ionic radius than Se, the substitution should produce positive chemical pressure. Recently, J. Guo *et al.* reported that both the $T_c^{onset}$ and $T_c^{zero}$ of $K_{0.8}Fe_{1.7}Se_2$ decreased with increasing pressure and superconductivity disappeared under pressure above 9.2 GPa.[29] Similar suppression effect of superconductivity under external pressure was also found in $K_{0.85}Fe_2Se_{1.80}$ and $Cs_{0.86}Fe_{1.66}Se_2$.[30] However, Kawasaki *et al.* reported that $T_c^{onset}$ increases with increasing pressure, while $T_c^{zero}$ decreases.[31] In our work, doping with S does not decrease the transition temperature, which suggests that the effect of chemical pressure on superconductivity is different from that of external pressure. Considering $AFe_2Se_2$, the bond between the FeSe layers is via weak van-der-Waals force, which is fragile to the application of external pressure. In contrast, the substitution of S for Se may have more effect on modulating the bond between Fe-Se, thereby decrease the anion height from Fe-layer. In FeTe(Se), the anion height is believed to play a key factor for determining the superconductivity and magnetism.[28]

We further performed ESR measurements to study the micro-magnetism of the $K_{0.8}Fe_2Se_{1.4}S_{0.4}$ single crystal. The benefit of ESR is that it can give dynamic information of the local moment and magnetic correlation. In cuprate superconductors and their parent compounds, ESR has been shown to be a highly sensitive tool to study the spin fluctuations and magnetic interactions. For the iron-based superconductors, ESR experiments on $LaFeAsO_{1-x}F_x$ and $EuFe_2As_2$-related systems as well as FeSe(Te) have also been carried out recently, which showed clear evidence for the existence of local moment and spin fluctuations.[32-36]

Figure 4 shows the ESR spectra of the $K_{0.8}Fe_2Se_{1.4}S_{0.4}$ sample from 2 K to 300 K. As the spectra signal at the temperature range of 40 K~300 K is very weak, we multiply the intensity by 100. At room temperature, there is a weak paramagnetic signal locating at about 3200 Oe. The resonance field ($H_{res}$) is defined as the magnetic field corresponding to the midpoint between the highest and lowest points in the ESR spectrum. In traditional metal, no obvious paramagnetic resonance signal is expected due to rapid spin-lattice relaxation. The observed resonance signal here should be considered as arising from the local moment of Fe ions, similarly to the case of $LaFeAsO_{1-x}F_x$.[32] With decreasing temperature from 300 K, the spectrum broadens and the resonance field moves to low magnetic fields followed by abrupt decrease of the resonance intensity below 120 K. The strong temperature dependence of resonance field and spectrum width is in agreement with that observed in other iron-based superconductors and can be explained as the enhancement of spin fluctuation. At temperatures between 40 - 70 K, no evident resonance signal coming from the local moment is observed, indicating that spin fluctuation enhances with decreasing temperature towards $T_c$. We notice that the temperature corresponding to the abrupt decrease of spectrum intensity is close to the hump temperature in resistivity curve, which might be hint for the correlation between the hump and spin fluctuation. This is in consistent with the pressure effect on the hump and superconductivity, which showed that high pressure suppresses the hump and superconductivity simultaneously.[29]

With further decreasing temperature below 35 K, a diplike signal appears at low fields typical for the magnetic shielding feature, indicating the appearance of superconductivity. Upon cooling, the intensity of the diplike signal increases. It is interesting to note that a novel resonance spectrum develops below 20 K, which is unexpected in conventional superconductors. Its resonance field shifts to higher field and the intensity of the new spectrum increases gradually with decreasing temperature. The signal cannot be attributed to the penetration of magnetic flux in the Shubnikov phase, because it only contributes to broad non-resonant microwave absorption.

To understand the novel resonance spectra at temperatures below $T_c$, two possibilities should be considered. One is that the signal comes from the local moment of Fe ions, possessing the same origin as that displayed at room temperature. As discussed above, the signal is suppressed by the enhancement of spin fluctuation. Upon cooling below $T_c$, the spin fluctuation gradually weakens and reappearance of the resonance signal can be expected. The existence of local moment in the temperature range from below $T_c$ to 300 K has also been reported in superconducting FeSe$_{0.35}$Te$_{0.65}$.[37] Inelastic neutron scattering performed by Xu *et al.* revealed a substantial moment (0.26 µB/Fe) that showed little change upon cooling to superconducting state.[37] The other possibility is the magnetic resonance spectrum at low temperatures corresponds to a novel magnetic state, which evolves in the superconducting matrix. Nevertheless, the unexpected ESR resonance spectrum below the superconducting transition temperature gives direct evidence for the coexistence of superconductivity and magnetism in K$_{0.8}$Fe$_2$Se$_{1.4}$S$_{0.4}$ crystal.

In summary, we successfully grew single crystals of a new superconductor K$_{0.8}$Fe$_2$Se$_{1.4}$S$_{0.4}$ with the superconducting transition temperatures $T_c^{onset}$ = 32.8 K. The ESR data shows clear evidence for the existence of spin fluctuation which enhances upon cooling towards $T_c$. At temperatures below $T_c$, a novel resonance signal develops gradually accompanied by the superconducting diamagnetic signal, indicating the coexistence of superconductivity with magnetism in K$_{0.8}$Fe$_2$Se$_{1.4}$S$_{0.4}$ crystal.

This research was financially supported by the National Key Basic Research of China Grant, Nos. 2007CB925001, 2010CB923403, and 2011CBA00111, and the National Nature Science Foundation of China Grant 11074258.

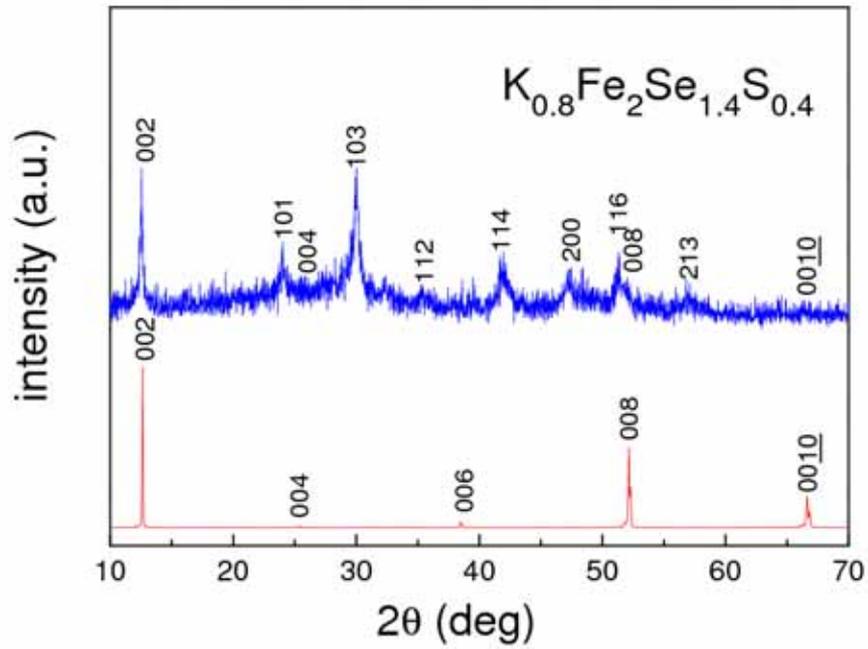

Figure 1　Powder x-ray diffraction pattern and single crystal XRD pattern for $K_{0.8}Fe_2Se_{1.4}S_{0.4}$.

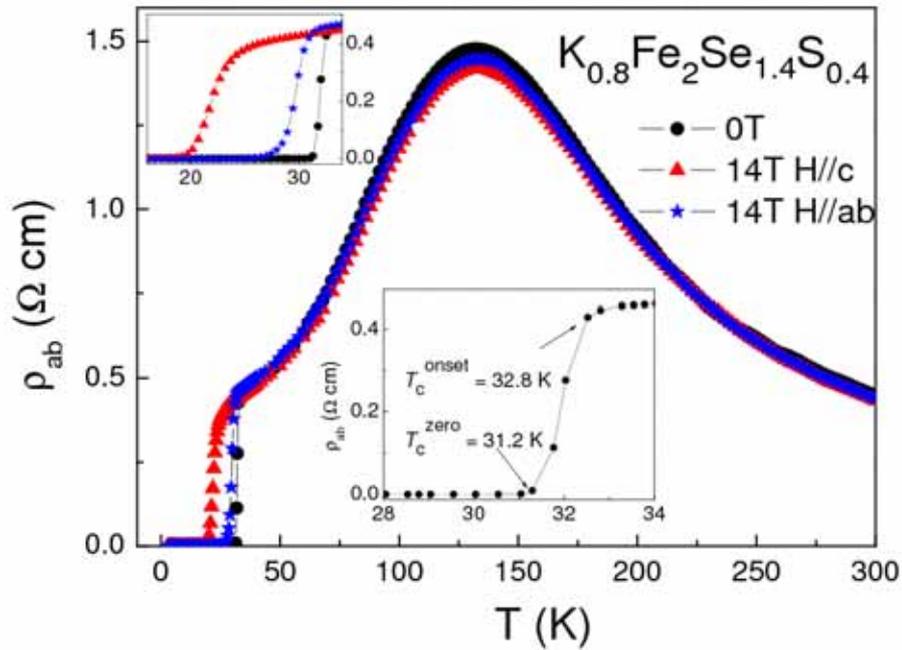

Figure 2　In plane resistivity as a function of temperature for $K_{0.8}Fe_2Se_{1.4}S_{0.4}$ crystal with the magnetic field of 14 T parallel to the *c*-axis and the *ab*-plane respectively. The lower inset shows enlarge views of superconducting transition from 28 to 34 K. The upper inset shows the zoom plot of resistivity around superconducting transition under external magnetic field.

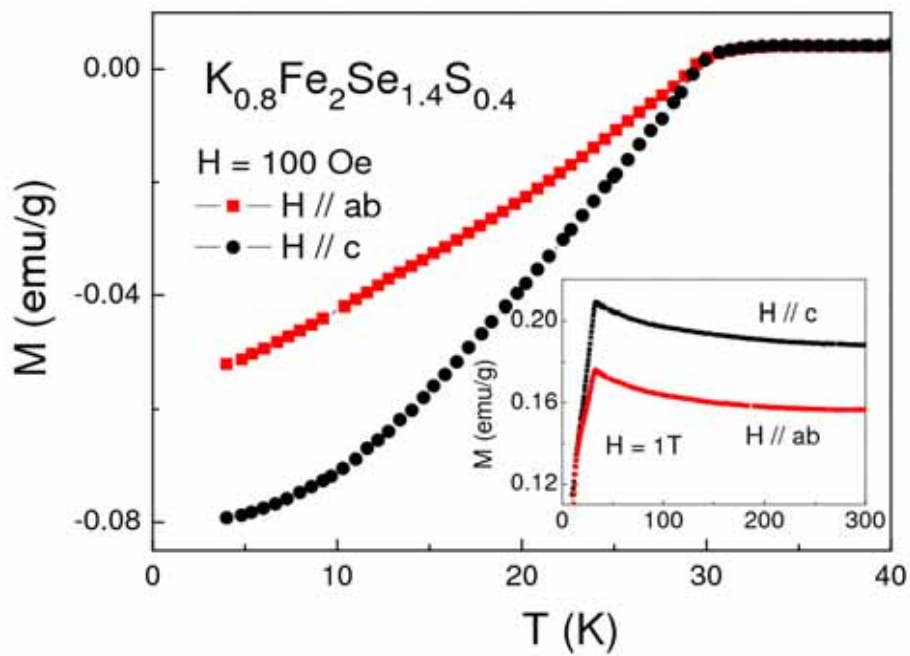

Figure 3  The magnetization at 100 Oe for single crystal $K_{0.8}Fe_2Se_{1.4}S_{0.4}$ with the magnetic field parallel and perpendicular to ab-plane, respectively. The inset shows the magnetic susceptibility under a magnetic field of 1 T.

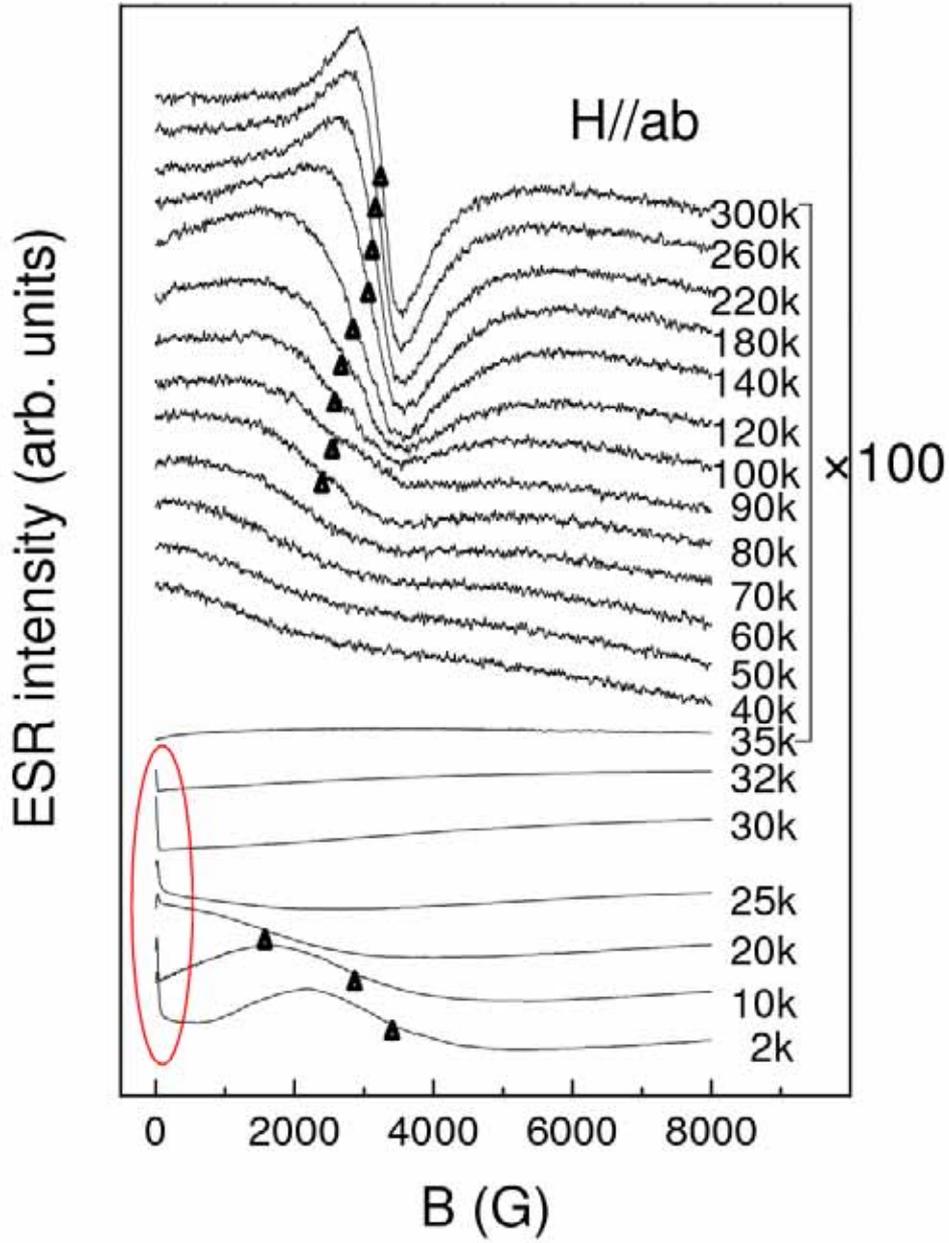

Figure 4 Electron spin resonance (ESR) spectra at different temperatures for $K_{0.8}Fe_2Se_{1.4}S_{0.4}$ crystal with sweeping magnetic field along ab-plane. The superconducting feature spectrum is marked by a red

circle. Solid triangle denotes the resonance field.